# Wafer-Scale Integration of Graphene-Based Photonic Devices

Marco A. Giambra,[†] Vaidotas Mišeikis,[†] Sergio Pezzini, Simone Marconi, Alberto Montanaro, Filippo Fabbri, Vito Sorianello, Andrea C. Ferrari, Camilla Coletti,*,[‡] and Marco Romagnoli*,[‡]

ACCESS　|　Metrics & More　|　Article Recommendations

**ABSTRACT:** Graphene and related materials can lead to disruptive advances in next-generation photonics and optoelectronics. The challenge is to devise growth, transfer and fabrication protocols providing high ($\geq 5000$ cm$^2$ V$^{-1}$ s$^{-1}$) mobility devices with reliable performance at the wafer scale. Here, we present a flow for the integration of graphene in photonics circuits. This relies on chemical vapor deposition (CVD) of single layer graphene (SLG) matrices comprising up to ∼12000 individual single crystals, grown to match the geometrical configuration of the devices in the photonic circuit. This is followed by a transfer approach which guarantees coverage over ∼80% of the device area, and integrity for up to 150 mm wafers, with room temperature mobility ∼5000 cm$^2$ V$^{-1}$ s$^{-1}$. We use this process flow to demonstrate double SLG electro-absorption modulators with modulation efficiency ∼0.25, 0.45, 0.75, 1 dB V$^{-1}$ for device lengths ∼30, 60, 90, 120 $\mu$m. The data rate is up to 20 Gbps. Encapsulation with single-layer hexagonal boron nitride (hBN) is used to protect SLG during plasma-enhanced CVD of Si$_3$N$_4$, ensuring reproducible device performance. The processes are compatible with full automation. This paves the way for large scale production of graphene-based photonic devices.

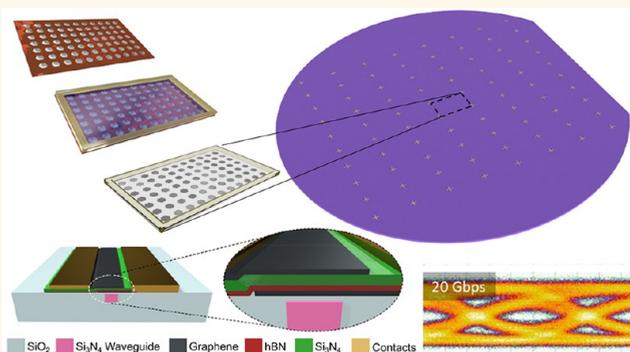

## INTRODUCTION

Graphene is ideally suited for photonics and optoelectronics,[1−4] in particular, for optical[5] and data communications,[2,5,6] including virtual Internet servers and data centers.[2] In 2020, the global IP data traffic, mostly through cloud and data centers, was in the range of several zettabytes (ZB),[7] *i.e.*, >10$^{21}$ bytes exchanged in one year. The connection of an ever-increasing number of people and things to the Internet (Internet of things, IoT[8]) is pushing the requirements in terms of bandwidth (BW), defined as amount of data exchanged per unit time,[9] and the energy consumed by a device to exchange one bit of information.[10] By 2023, >27 billion devices are expected to be connected.[7] COVID-19 has forced people to stay at home, working and learning remotely as never before.[11] This resulted in an increase by 20−100% of the fixed residential network[11] and 10−20% change in traffic levels on the mobile network.[11] Thus, there is a renewed demand of traffic for applications, such as teleconferencing, video streaming, and online games.[12] Photonic technologies play a key role to satisfy these requirements. Photonic devices for next-generation telecom and datacom networks require >100 Gbps BW per single lane,[13] a small footprint (<mm$^2$),[14] a low loss of optical power within the device due to optical coupling (<1 dB),[15] propagation loss <2 dB cm$^{-1}$,[16] insertion loss (IL), *i.e.* power loss due to insertion of a device,[17] <5 dB,[18,19] low energy cost <1 mW/GHz or, equivalently, <1 pJ/bit,[20,21] and low cost of manufacturing (<$10/Gbps in 2020,[2] decreasing to <$1/Gbps by 2025).[22] For these reasons, photonic devices based on alternatives to the established silicon on insulator, SOI,[23] and InP technologies are being investigated.[24] Silicon photonics (SiPh) modulators for ≥30 Gbaud applications have IL ∼ 2−3 dB higher than InP- and LiNbO$_3$-based modulators,[25] because of the free carrier effect,[26] requiring device lengths in the mm scale. The baud represents the data in a transmission channel. It is a symbol that contains a string of "n" bits.[27] Typically, in optical communication systems "n" is 1 to 6.[27] The bit rate is defined as baud rate times "n".[17,27]

More compact and energy efficient devices were demonstrated exploiting resonant structures, e.g., microring resonators,[28] or the Franz−Keldysh effect in Si−Ge alloys.[29] However, these have intrinsic wavelength selectivity.[29] InP technology provides modulators with size similar to SiPh,[30] large BW (>50 GHz),[31] but with a higher cost of manufacturing,[31] due to the greater cost of InP wafers with respect to Si ones.[14,32,33]

Graphene-based photonics is very promising, as graphene is fully compatible with SiPh,[2] it has electro-absorption[2,34] and electro-refraction properties,[2,34] and it can be used for light modulation[2] and photodetection.[1,3] The linear gapless energy-momentum relation of the massless Dirac Fermions in single-layer graphene (SLG) leads to high mobility at room temperature (RT) ($\mu$ > 100000 cm$^2$ V$^{-1}$ s$^{-1}$)[35−40] and pronounced (more than 1 order of magnitude)[35−38] ambipolar electric field effect,[41] such that the surface conductivity, $\sigma$, can be tuned by applying a gate voltage.[41] The tuning of $\sigma$ influences the optoelectronic properties of SLG.[42,43] $\sigma$ is a complex quantity, affecting both absorption and refraction of light interacting with SLG.[42] When SLG is placed on a waveguide (WG) core, the guided light interacts with SLG, allowing a much larger absorption with respect to normal incidence.[44] The absorption coefficient for SLG on a SOI WG is up to 0.1 dB $\mu$m$^{-1}$,[45] depending on SLG doping[45] and distance from the WG core center.[46]

SLG has been used for electron absorption[46,47] and electron refraction modulation,[48,49] switching,[50] and photodetection.[1−3,51−55] Reference 46 reported electron absorption modulators (EAMs) based on SLG transferred on a 7 nm Al$_2$O$_3$ layer deposited on a Si WG. This configuration was improved by using a SLG-insulator-SLG stack, i.e., a double SLG (DSLG),[2] on an undoped Si WG.[45,47,56] This has two main advantages: (1) the use of a passive WG platform, i.e., pure dielectric WGs, without implantation or epitaxy processes typically employed in SiPh[57,58] or InP,[31] simplifying the manufacturing process, with a consequent cost reduction; (2) enhanced modulation due to the interaction of two SLGs with the WG mode.[34] Single-mode WGs have typical dimensions which depend on the refractive index of the guiding material.[59] SiPh single-mode WGs, guiding only the fundamental mode,[59] have a typical width ∼480 nm when realized on 220 nm SOI.[60] Si$_3$N$_4$ single-mode WGs have larger width ∼1 $\mu$m, depending on Si$_3$N$_4$ thickness,[39] because of the lower refractive index ($n$ = 1.98 for Si$_3$N$_4$[39] compared to 3.47 for Si at 1550 nm).[61] The larger width of Si$_3$N$_4$ WGs helps simplify the technology because it requires less stringent lithography resolution and also reduces costs, making small (∼10000 pieces/year) and medium (∼100000−1000000 pieces/year) production volumes more affordable than in SOI or InP manufacturing lines.[31] This means that the volume (i.e., number of chips) threshold to implement a product in a Si fab can be reduced by using Si$_3$N$_4$. This enables the cost-effectiveness of medium-volume products (∼10000−100000 chips per year),[2] thus opening medium-volume markets (e.g., long haul telecom systems).[2]

To reach a high technology-readiness level (TRL > 8, i.e., system complete and qualified),[62] adequate for photonic device production, scalable techniques for SLG growth and transfer are needed. Chemical vapor deposition (CVD) on Cu yields SLG that, when encapsulated in hexagonal boron nitride (hBN), has electronic and structural quality (defect density, scattering time, and $\mu$) comparable to exfoliated SLG.[35,37,38,63] There has been significant progress for SLG scalable growth on dielectrics, such as SiO$_2$[64] and Al$_2$O$_3$,[65] and on CMOS-compatible Ge,[66−68] but with RT $\mu$ limited to ∼2000 cm$^2$ V$^{-1}$ s$^{-1}$.[65] Hence, as of 2020, the most common approach to obtain $\mu$ > 5000 cm$^2$ V$^{-1}$ s$^{-1}$ is to transfer SLG grown on Cu to the target substrate.[69] The so-called "wet" transfer[70,71] typically involves chemical etching Cu to release SLG.[69,72] Alternatively, SLG can be released from the growth substrate electrochemically[73,74] or by oxidizing Cu at the SLG interface.[75] The released SLG is then directly picked up from the aqueous solution using the target wafer, with alignment accuracy ≥1 $\mu$m.[76] Wet-transferred SLG has $\mu$ ∼ 10$^3$ cm$^2$ V$^{-1}$ s$^{-1}$,[69] which can be improved by 2 orders of magnitude by hBN encapsulation.[37] "Fully dry" transfer[35] is based on direct pick-up of SLG from Cu using exfoliated flakes of hBN or other layered materials (LMs), such as WSe$_2$.[39] In this approach, SLG is released from Cu and encapsulated without contact with water or solvents,[35] resulting in $\mu$ > 3 × 10$^5$ cm$^2$ V$^{-1}$ s$^{-1}$ at RT.[39] Thus far, scalability is limited by the size of exfoliated hBN flakes (up to ∼100 $\mu$m),[77] but CVD hBN or amorphous BN could be used in future to solve this. The "semi-dry" approach consists in SLG delamination from Cu in an aqueous solution either electrochemically[76] or by Cu oxidation,[78] followed by lamination on the target substrate in dry conditions. This yields $\mu$ as high as in "fully dry" transfer after hBN encapsulation[38] while allowing scalability.[76]

Here, we implement an aligned semidry transfer of SLG, based on electrochemical delamination in NaOH, and subsequent handling of a suspended polymer/SLG membrane using a frame. This approach avoids the contact of the target substrate with the aqueous solution and allows deterministic placement of SLG single crystals (SC) with ∼1 $\mu$m precision in the X and Y plane, thanks to a transfer setup equipped with micrometric actuators. We use a freestanding carrier membrane, comprising 2 polymer layers. This enables semidry transfer of large SLG matrices (up to ∼12000 SLG-SCs) with coverage >80% of the target photonics device area, and integrity in terms of SLG continuity.

We report wafer-scale fabrication of DSLG EAMs on Si$_3$N$_4$ WGs based on a stack of two SLGs separated by ∼17 nm Si$_3$N$_4$. We report 30 EAMs, on 4 chips from the same wafer, with uniform performance ±10%, demonstrating wafer-scale scalability and reproducibility of the complete process. We use monolayer (1L) CVD-hBN for SLG encapsulation, to protect SLG during Si$_3$N$_4$ deposition by plasma-enhanced CVD (PECVD). We get a contact resistance ∼500 Ω $\mu$m for $E_F$ > 0.2 eV, allowing us to achieve a cutoff frequency, i.e., the frequency at which energy flowing through the system is reduced rather than passing through,[17] ∼4 GHz for 120 $\mu$m EAMs, and ∼12 GHz for 30 $\mu$m ones. The operation speed is ∼20 Gbps, the highest to date in Si$_3$N$_4$ without using resonating devices. Higher speeds have only been demonstrated in Si$_3$N$_4$ with resonating devices. For example, SLG on Si$_3$N$_4$ modulators working up to 22 Gbps were reported on microring resonators,[56] while up to 40 Gbps was demonstrated by using piezoelectric lead zirconate titanate (PZT) thin films on Si$_3$N$_4$ microring resonators.[80] Because of the gapless nature of SLG,[1−3,81] SLG photonics can operate at any wavelength, unlike refs 56 and 80, which were limited to the specific resonant wavelength.

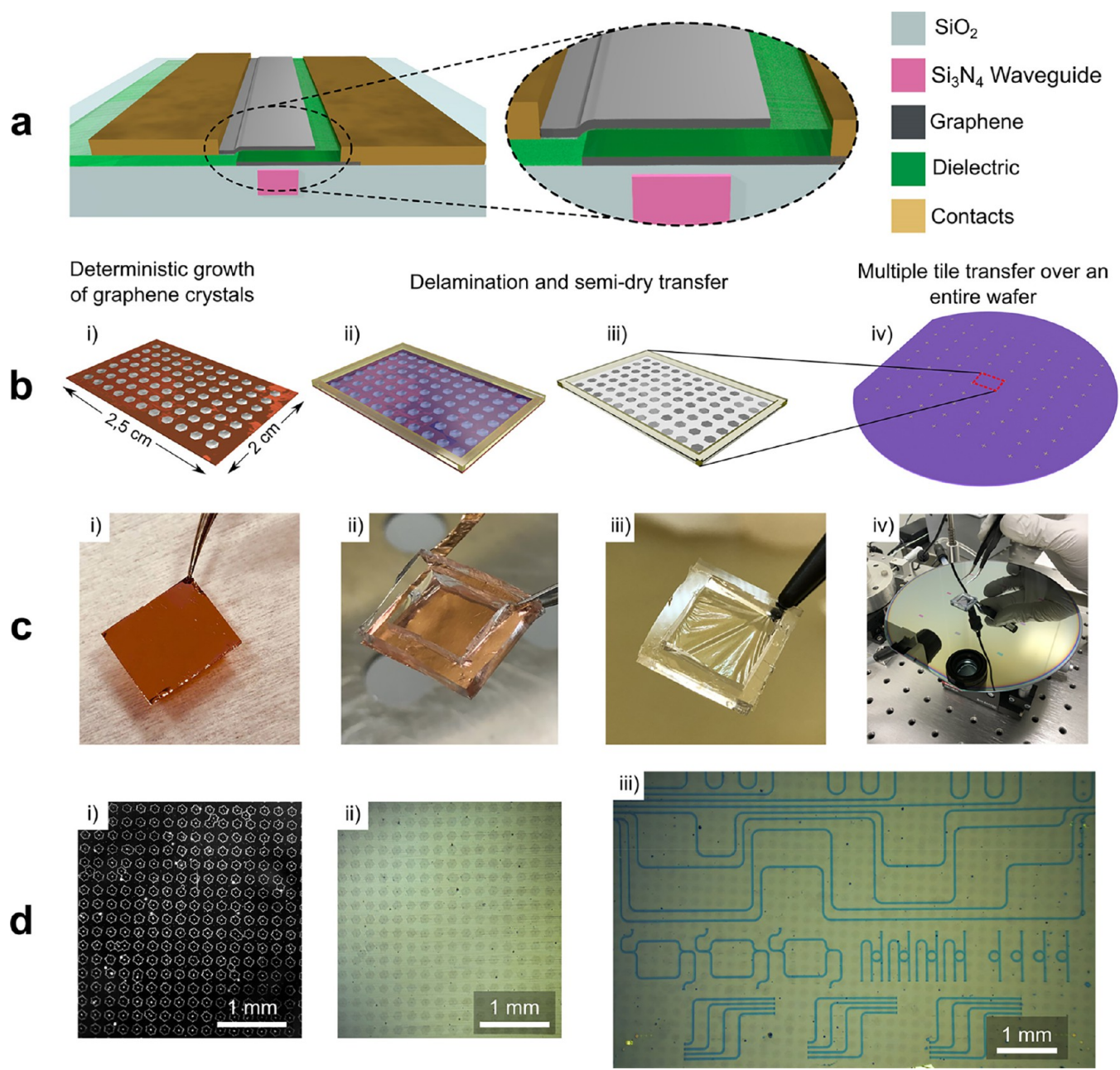

Figure 1. (a) Schematic cross-section of DSLG EAM. (b) Multiple tile stamping: (i) schematic of SC-SLG matrix on Cu, (ii) SC-SLG matrix on Cu covered with freestanding membrane and a frame, enabling aligned transfer, (iii) delaminated SC-SLG matrix with freestanding membrane and frame, (iv) transferred SLG on target wafer. (c) Photos of: (i) as-grown SLG on Cu, (ii) Cu with PDMS frame attached, (iv) suspended polymer/SLG membrane and 150 mm photonic wafer with laminated SLG. (d) Optical micrographs of (i) SC-SLG on Cu by dark field imaging, (ii) suspended SLG-SCs on polymer membrane and (iii) transferred SLG SC matrix on target wafer with photonic circuits.

## RESULTS AND DISCUSSION

Our DSLG EAMs comprise two SLGs on a passive $Si_3N_4$ WG, separated by a ~17 nm $Si_3N_4$ dielectric, Figure 1a. Three factors ensure scalable fabrication with reproducibility: (i) wafer-scale source material with crystal size comparable to that of single devices, to avoid grain boundaries; (ii) semidry transfer with low impact on SLG cleanliness and electrical properties; (iii) SLG protection prior and during dielectric deposition. In ref 76, we addressed (i) by preparing SLG SC matrices. This approach is compatible with the requirements of integrated photonics, allowing tailored growth of SLG according to the geometry of the photonic circuits. The lateral dimensions of the SLG SCs can be tuned from tens to hundreds of micrometers.[45,56,82] Deterministic growth relies on pretreating Cu by electropolishing, to reduce surface contaminations and improve surface flatness. Cu is then patterned with 5 $\mu$m Cr seeds at the desired SLG crystal locations. This is done by using optical lithography and thermal evaporation of 25 nm Cr. The growth is performed in a cold-wall CVD reactor (Aixtron BM Pro) at 1060 °C by using Ar annealing to maintain a low nucleation density (~10 crystals per mm$^2$).[79] Due to residual oxidation in Cu, SLG nucleation requires surface impurities,[83] ensuring that SLG SCs nucleate only at the Cr seeds locations. The matrices of SLG SCs grown on Cu need to be released from the growth substrate and transferred to the target wafer (*e.g.*, a wafer containing WGs). To do so, we adapt our semidry transfer procedure[76] and build a dedicated transfer tool. To facilitate handling, SLG is coated with a polymer carrier membrane, and a semirigid polydimethylsiloxane (PDMS) frame is attached to the Cu foil perimeter. The transfer itself consists of two stages: (1) wet SLG electrochemical delamination from the growth substrate and (2) dry SLG aligned lamination on the target substrate. After the SLG electrochemical release from Cu in NaOH (see the Methods for details), the SLG/polymer membrane is rinsed several times in deionized (DI) water and dried in air. The freestanding membrane is supported by the PDMS frame and can be handled in dry conditions. The SLG SCs are attached to the membrane holder of the lamination tool, which allows angle adjustment with ~0.1° precision of the membrane with respect to the target wafer. The latter is brought in close proximity (~500 $\mu$m) to the membrane using a 4-axis micrometrical stage (X, Y, Z translation and $\Theta$ rotation). After aligning the SLG-SCs to the photonic structures, the target wafer is heated to ~100 °C and brought into contact with SLG, resulting in adhesion with the target photonics chip over the whole membrane. The alignment is performed using a 12× zoom microscope lens attached to a Digital single-lens reflex (DSLR) camera. The PDMS frame is then detached from the sample, and placed in acetone for the polymer removal.

During the delamination of SLG from Cu and alignment to the target substrate, the freestanding polymer-SLG membrane is supported by a semirigid frame attached to the perimeter of the sample, Figure 1c. In ref 76, the frame was made from polyimide (Kapton) tape and bonded to the sample using an adhesive, with the risk of chemical reaction with the NaOH electrolyte contaminating the transferred SLG. To mitigate this, here we use PDMS-based support frames, which can be bonded to flat surfaces without any adhesive, thus ensuring transfer cleanliness. An alternative could be to use a solid PDMS stamp,[84] which may also handle SLG. However, PDMS is not compatible with the lamination temperature (105 °C), due to its large (~3.1 × 10$^{-4}$ K$^{-1}$) thermal expansion coefficient.[85] SLG-SCs attached to a PDMS stamps can develop nanometer-sized cracks when heated to 100 °C. Our method also relies on a bilayer carrier polymer comprising 1.5 $\mu$m poly(propylene carbonate) (PPC) and 100 nm PMMA, instead of the PMMA support of ref 76. The different glass transition temperatures, $T_G$, of PPC (37 °C)[86] and PMMA (105 °C)[86] allow us to have a membrane with variable mechanical properties, which can be controlled with $T$. At ambient $T$, during delamination and SLG SC alignment, both polymers are kept <$T_G$, thus providing a rigid support to the freestanding membrane and preventing SLG damage. When the SLG SCs are aligned to the required position on the target wafer, SLG can be laminated on the substrate by heating to ~100 °C, well above the PPC $T_G$. The relatively thick and viscous PPC layer compared to PMMA allows the membrane to attach to the wafer and conform to surface structures, such as metal contacts, while retaining the integrity due to the solid, yet thin (~100 nm), PMMA layer, still below the PMMA $T_G$. Crucially, in the lamination stage, the target substrate does not come into contact with an aqueous solution. Therefore, the transfer can be repeated on different areas of the same wafer, without risk of SLG delamination or increased contamination.

This enables the growth of SLG on a smaller scale, with greater control of strain and doping, than that currently achievable[87,88] when performing growth and transfer on full 150 or 200 mm wafers. The target wafer can then be populated via several transfers, as shown schematically in Figure 1b. Before each SLG transfer, photonics WGs are prepared by rinsing the chip in acetone and 2-propanol, followed by a deep cleaning in a resist remover (AR-600 71) for 2 min. Following SLG transfer on the WGs, the fabrication of the DSLG stack is performed as follows. SLG is patterned and etched using electron-beam lithography (EBL) (Zeiss Ultra Plus) and reactive ion etching (RIE) (Sistec). The bottom SLG contacts are deposited via thermal evaporation of Ni/Au. A protective 1L-hBN film is transferred over the whole chip area using the semidry procedure described above.

A 17 nm Si$_3$N$_4$ gate dielectric is deposited over the whole area. Si$_3$N$_4$ is chosen over other dielectrics, such as Al$_2$O$_3$, HfO$_2$, or hBN, due its high breakdown field (>10 MV cm$^{-1}$).[89] PECVD can be used to deposit uniform Si$_3$N$_4$ with thickness <20 nm and root mean square (RMS) roughness <0.5 nm.[89] Top SLG SCs are then placed using aligned semidry transfer. The top structure of the modulator is fabricated using identical methods to the bottom layer (see the Methods for details).

The SLG crystals are characterized throughout the fabrication process by Raman spectroscopy with a Renishaw InVia at 532 nm, laser power ~1 mW, and acquisition time ~4s. The laser spot size is ~0.8 $\mu$m, as determined by the razor blade technique.[90,91] We present a detailed step-by-step procedure to acquire and analyze Raman spectra throughout the fabrication of wafer scale SLG-based devices. This ensures quality control as well as reproducibility. The complete set of data we provide enables independent assessment of our results. Tables 1 and 2 present a summary of the Raman fitting parameters and corresponding defect density, Fermi level ($E_F$), and strain.

Table 1. Raman Fit Parameters from Figure 2d–g and Corresponding Defect Density, $E_F$, and Strain

| | SLG on SiO$_2$ | SLG after Si$_3$N$_4$ deposition with hBN encapsulation | SLG after Si$_3$N$_4$ deposition without hBN encapsulation |
|---|---|---|---|
| Pos(G) (cm$^{-1}$) | 1585.5 ± 0.7 | 1590.3 ± 1.5 | 1590 ± 1.6 |
| FWHM(G) (cm$^{-1}$) | 10.5 ± 1.0 | 11.8 ± 1.7 | 12.0 ± 1.9 |
| Pos(2D) (cm$^{-1}$) | 2678 ± 1.2 | 2684.6 ± 1.8 | 2679.3 ± 1.8 |
| FWHM(2D) (cm$^{-1}$) | 26.9 ± 0.8 | 32.5 ± 1.5 | 33.8 ± 2 |
| I(2D)/I(G) | 2.6 ± 0.3 | 2.3 ± 0.3 | 1.8 ± 0.2 |
| A(2D)/A(G) | 6.8 ± 0.6 | 6.5 ± 0.7 | 5.1 ± 0.5 |
| I(D)/I(G) | <0.02 | <0.05 | 0.48 ± 0.06 |
| defect density (10$^{11}$ cm$^{-2}$) | <0.05 | <0.10 | 1.98 ± 0.3 |
| $E_F$ (meV) | 190 ± 30 | 220 ± 40 | 300 ± 40 |
| uniaxial strain (%) | −0.08 ± 0.08 | 0.06 ± 0.12 | −0.14 ± 0.12 |
| (biaxial strain) (%) | (−0.03 ± 0.03) | (0.02 ± 0.04) | (−0.06 ± 0.05) |

Figure 2b shows representative spectra of SLG on 285 nm SiO$_2$/Si, before (black) and after Si$_3$N$_4$ deposition, with (orange) and without (dark cyan) capping of SLG with 1L-hBN (see sketch in Figure 2a). The Raman signature of 1L-hBN is weak indicating the low quality of the commercial 1L-hBN.[92] The transferred SLG spectrum has a 2D peak with a single Lorentzian shape and with a full width at half-maximum FWHM (2D) ∼ 26.7 cm$^{-1}$, a signature of SLG.[93] The G peak position, Pos(G), is ∼1583.7 cm$^{-1}$, with FWHM(G) ∼ 12.4 cm$^{-1}$. The 2D peak position, Pos(2D) is ∼2676 cm$^{-1}$, while the 2D to G peak intensity and area ratios, $I(2D)/I(G)$ and $A(2D)/A(G)$, are ∼3.1 and ∼6.8, respectively. No D peak is observed, indicating negligible defects concentration.[94,95] After Si$_3$N$_4$ deposition, the Raman spectrum of exposed SLG (i.e., without 1L-hBN capping) has Pos(G) ∼ 1590 cm$^{-1}$, FWHM(G) ∼ 12 cm$^{-1}$, Pos(2D) ∼ 2679 cm$^{-1}$, FWHM(2D) ∼ 33.8 cm$^{-1}$, $I(2D)/I(G)$ ∼ 1.8, $A(2D)/A(G)$ ∼ 5.1, and $I(D)/I(G)$ ∼ 0.5. The latter indicates the creation of Raman active defects, which also act as scattering centers for the charge carriers[96,97] (1 order of magnitude $\mu$ decrease was reported in ref 96 when going from $I(D)/I(G)$ ∼ 0.01 to ∼ 0.5). Carrier scattering limits the performance of SLG EAMs, in terms of modulation efficiency (slope of the transmission variation as a function of applied voltage[17]) and maximum extinction ratio (ER)[17] (i.e., the ratio between maximum and minimum of light transmission[17]). The effect of defects on FWHM(G),[95] which remains almost unchanged after Si$_3$N$_4$ deposition, is likely compensated by the increased doping.[98] The Raman data indicate that $E_F$ of SLG after transfer is ∼170 meV (hole doping).[99,100] $E_F$ in the exposed SLG increases to ∼290 meV.[99,100]

Table 2. Raman Fit Parameters from Figure 3b−e and Corresponding Defect Density, $E_F$, and Strain

|  | SLG on SiO$_2$ | SLG after hBN encapsulation | SLG after Si$_3$N$_4$ deposition | SLG on Si$_3$N$_4$ |
| --- | --- | --- | --- | --- |
| Pos(G) (cm$^{-1}$) | 1583.1 ± 0.5 | 1584 ± 0.9 | 1593.8 ± 1.4 | 1582.3 ± 0.7 |
| FWHM(G) (cm$^{-1}$) | 11 ± 1.1 | 12.2 ± 1.5 | 8.8 ± 1.9 | 14 ± 1.2 |
| Pos(2D) (cm$^{-1}$) | 2675.2 ± 0.6 | 2678.5 ± 1.7 | 2687.1 ± 2.6 | 2674.1 ± 0.9 |
| FWHM(2D) (cm$^{-1}$) | 23 ± 0.8 | 25.9 ± 1.3 | 32.7 ± 2.6 | 23.4 ± 1 |
| $I(2D)/I(G)$ | 4.3 ± 0.4 | 3.9 ± 0.4 | 1.8 ± 0.3 | 4.6 ± 0.5 |
| $A(2D)/A(G)$ | 8.9 ± 0.7 | 8.4 ± 0.8 | 6.8 ± 1.1 | 7.7 ± 0.7 |
| $I(D)/I(G)$ | <0.02 | <0.02 | 0.11 ± 0.10 | <0.05 |
| defect density (10$^{11}$ cm$^{-2}$) | <0.05 | <0.05 | 0.40 ± 0.4 | <0.10 |
| $E_F$ (meV) | <100 | <100 | 250 ± 50 | <100 |
| uniaxial strain (%) | 0.07 ± 0.02 | 0.11 ± 0.03 | 0.13 ± 0.13 | 0.03 ± 0.04 |
| (biaxial strain) (%) | (0.03 ± 0.01) | (0.04 ± 0.01) | (0.05 ± 0.05) | (0.01 ± 0.01) |

SLG capping with 1L-hBN is used to protect SLG during PECVD (at 350 °C) of Si$_3$N$_4$. The SLG spectra with hBN capping after Si$_3$N$_4$ deposition have Pos(G) ∼ 1590 cm$^{-1}$, FWHM(G) ∼ 11.8 cm$^{-1}$, Pos(2D) ∼ 2684 cm$^{-1}$, FWHM(2D) ∼ 32.5 cm$^{-1}$, $I(2D)/I(G)$ ∼ 2.3, $A(2D)/A(G)$ ∼ 6.5. Figure 2c is a statistical comparison of $I(D)/I(G)$ in 800 spectra from 2 SLG SCs with Si$_3$N$_4$ on top (400 spectra each), one protected by 1L-hBN (orange), the other exposed to PECVD (dark cyan). Ninety-eight percent of the spectra on hBN-encapsulated SLG have $I(D)/I(G)$ < 0.1. One hundred percent of the nonencapsulated SLG have $I(D)/I(G)$ > 0.1, with an average $I(D)/I(G)$ ∼ 0.48, corresponding to a defect concentration ∼1.98 × 10$^{11}$ cm$^{-2}$ (taking into account the finite doping ∼300 meV).[95,101] Hence, capping with 1L-hBN limits the creation of Raman active defects, therefore contributing to preserve $\mu$.[96,97] SLG SCs exposed to Si$_3$N$_4$ deposition present cracked areas with an average crack size ∼10 $\mu$m, as for the optical microscopy image in Figure 2c (right inset).

Raman mapping is performed at 1 $\mu$m steps, over an area ∼20 $\mu$m × 20 $\mu$m on SLG transferred onto SiO$_2$/Si, and after Si$_3$N$_4$ deposition, with and without 1L-hBN. Figure 2d−g plots Raman data extracted from the maps: Pos(2D), FWHM(2D), FWHM(G), $A(2D)/A(G)$, as a function of Pos(G). Pos(G) depends on both doping[99,100] and strain.[102] The average Raman parameters from Figure 2d−g are in Table 1, together with the corresponding estimates of defect density, $E_F$, and strain.

The Raman data indicate $E_F$ after transfer ∼190 meV (hole doping).[99,100] HBN capping, in addition to limiting the generation of Raman active defects, keeps $E_F$ close to that of transferred SLG (∼220 meV). $E_F$ in exposed SLG increases to ∼300 meV.[99,100]

The Grüneisen parameters[102] rule the change of Pos(2D) and Pos(G) in response to strain. The G and 2D peaks do (do not) split for increasing uniaxial (biaxial) strain.[94] At low (≲0.5%) strain the splitting cannot be resolved.[102,103] Figure 3d plots the correlation between Pos(2D) and Pos(G). Linear fits in Figure 3d give a slope $\Delta$Pos(2D)/$\Delta$Pos(G) ∼ 1.37, 0.85, 1.1 for SLG after transfer, after Si$_3$N$_4$ with hBN, and without hBN, respectively. The slopes indicate that both doping and strain variations are present. We cannot exclude the presence (or coexistence) of biaxial strain. For uniaxial (biaxial) strain, Pos(G) shifts by $\Delta$Pos(G)/$\Delta\epsilon$ ∼ 23(60) cm$^{-1}$/%.[102−104] For intrinsic SLG ($E_F$ < 100 meV), the unstrained, undoped Pos(G) is ∼1581.5 cm$^{-1}$.[93,105] Taking into account the shift in Pos(G) due to finite doping ($E_F$ ∼ 190, 220, 300 meV for the three cases), we estimate a mean uniaxial(biaxial) strain $\epsilon$ ∼ −0.08%(∼−0.03%) for the transferred SLG, and ∼0.06% (∼0.02%) and ∼−0.14% (∼−0.06%) for the hBN-capped and exposed SLG after Si$_3$N$_4$ deposition, respectively.

After 1L-hBN-capping and PECVD deposition of Si$_3$N$_4$, DSLGs are completed by transferring top-layer SLG arrays onto Si$_3$N$_4$ by semidry transfer. The use of identical deterministically grown SC matrices ensures that bottom and top SLG overlap over the entire wafer area, enabling wafer-scale fabrication.

The assembly of DSLG is monitored by Raman spectroscopy. We collect 8909 spectra on 48 crystals (24 bottom-layer and 24 top-layer) over four portions of a 150 mm wafer (p-doped Si with 285 nm SiO$_2$). Figure 3a plots representative spectra taken after the main assembly steps: (1) transfer of bottom SLG arrays on SiO$_2$/Si (black), (2) transfer of 1L-hBN (red), (3) deposition of Si$_3$N$_4$ (orange), and (4) transfer of top SLG on Si$_3$N$_4$ (purple). The SLG spectra after transfer on SiO$_2$ (bottom-layer, black) and on Si$_3$N$_4$(top-layer, purple) have a 2D peak with a single Lorentzian shape and FWHM(2D) ∼ 22.2 and 23.1 cm$^{-1}$, respectively. Pos(G) is ∼1583.1 cm$^{-1}$ for SLG on SiO$_2$ and ∼1582.1 cm$^{-1}$ for SLG on Si$_3$N$_4$, with FWHM(G) ∼ 10.6 and 14.5 cm$^{-1}$, respectively. Pos(2D) is

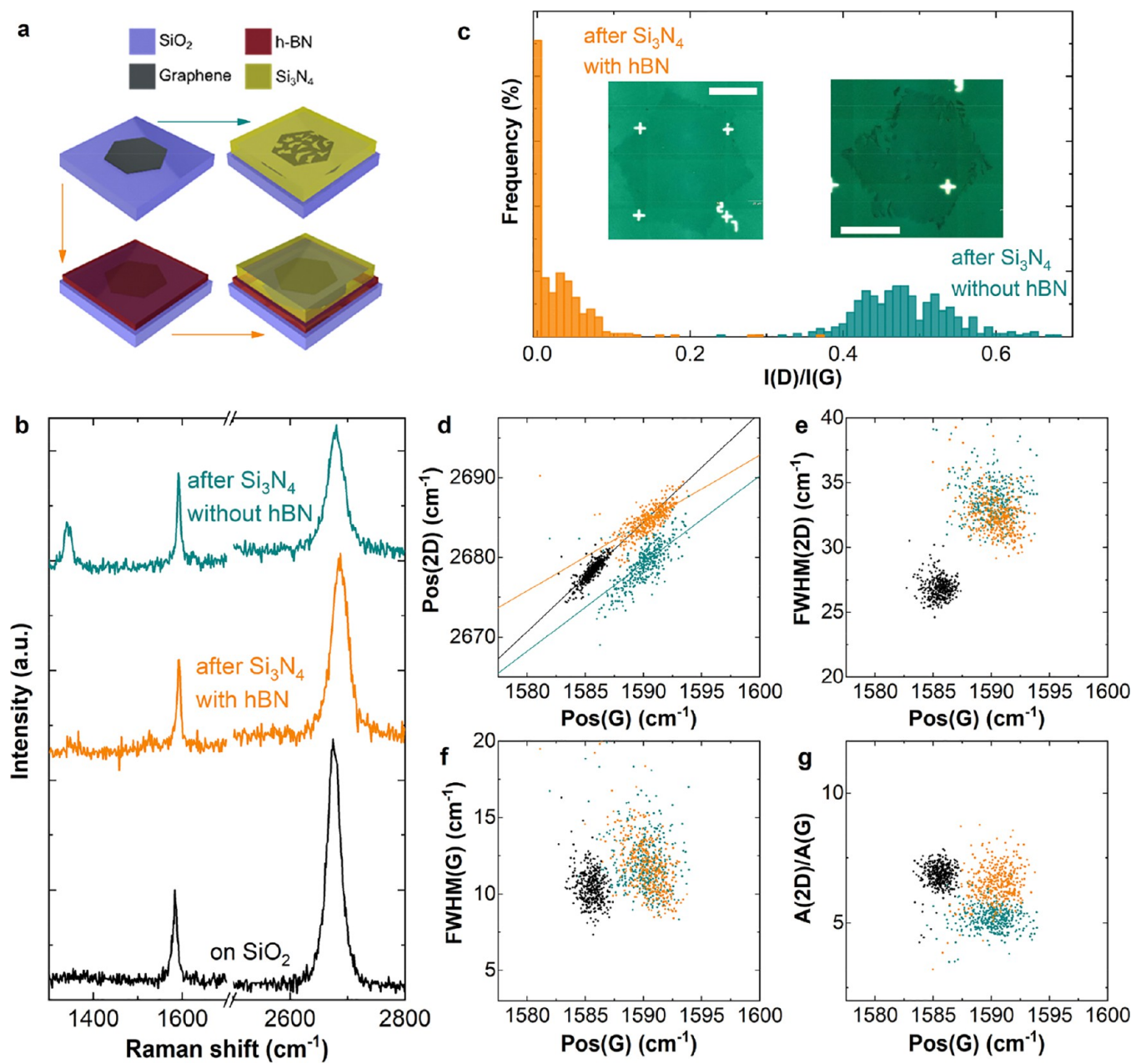

Figure 2. (a) Schematic representation of PECVD deposition of $Si_3N_4$ on SLG without (top, dark cyan arrow) and with (bottom, orange arrows) intermediate 1L-hBN. (b) Typical Raman spectra on SLG SCs after transfer (black) after $Si_3N_4$ PECVD, with (orange) and without (dark cyan) 1L-hBN. The same colors are used in the correlation plots d−g. (c) Distribution of $I(D)/I(G)$ from 800 spectra acquired on two SLG SCs, one protected (orange bars), the other exposed (dark cyan bars). Inset: optical micrographs of the two SCs, showing cracked areas in the exposed one. Scale bars 50 μm. (d) Pos(2D) as a function of Pos(G). Solid lines are linear fits of the data. (e) FWHM(2D) as a function of Pos(G). (f) FWHM(G) as a function of Pos(G). (g) A(2D)/A(G) as a function of Pos(G).

∼2675.3 and ∼2673.9 cm$^{-1}$, while $I(2D)/I(G)$ and $A(2D)/A(G)$, are ∼4.5 (on $SiO_2$), ∼5 (on $Si_3N_4$), ∼9.5 (on $SiO_2$), and ∼8 (on $Si_3N_4$). No D peak is observed, indicating negligible defect concentration.[94,95] The difference in FWHM(G) indicates reduced doping for the top-layer SLG on $Si_3N_4$.[99,106] The bottom-layer SLG spectra with hBN capping before (red) and after (orange) $Si_3N_4$ deposition have both a 2D peak with a single Lorentzian shape and FWHM(2D) ∼ 24.6 and 32.2 cm$^{-1}$. Pos(G) is ∼1583.9 and ∼1593.6 cm$^{-1}$ for hBN-capped SLG before and after $Si_3N_4$ deposition, with FWHM(G) ∼ 11.4 and 8.6 cm$^{-1}$, Pos(2D) ∼ 2678.7 and 2686.8 cm$^{-1}$. $I(2D)/I(G)$ and $A(2D)/A(G)$ are ∼4.1 and ∼8.9 before and ∼1.9 and ∼7.2 after the $Si_3N_4$ deposition. The shift of Pos(G) and decrease of FWHM(G), together with decrease of $I(2D)/I(G)$ and $A(2D)/A(G)$, indicate an increase in defect density, $E_F$, upon $Si_3N_4$ deposition.[99,100] In addition, a considerable (2500 cps at 1 mW power excitation) photoluminescence background is observed after $Si_3N_4$ deposition, which we attribute to the introduction of defects in 1L-hBN.[107] The broad band is peaked at ∼600 nm (inset, Figure 3a) similar to defect related broad emission in 1L-hBN.[107] Raman mapping is then performed on the SLG arrays at 10 μm steps. Figure 3b−e plots Pos(2D), FWHM(2D), FWHM(G), and A(2D)/A(G) as a function of Pos(G). Local variations in strain[98] and doping[98,99] produce a spread in Pos(G). The average Raman data of Figure 3b−e are presented in Table 2.

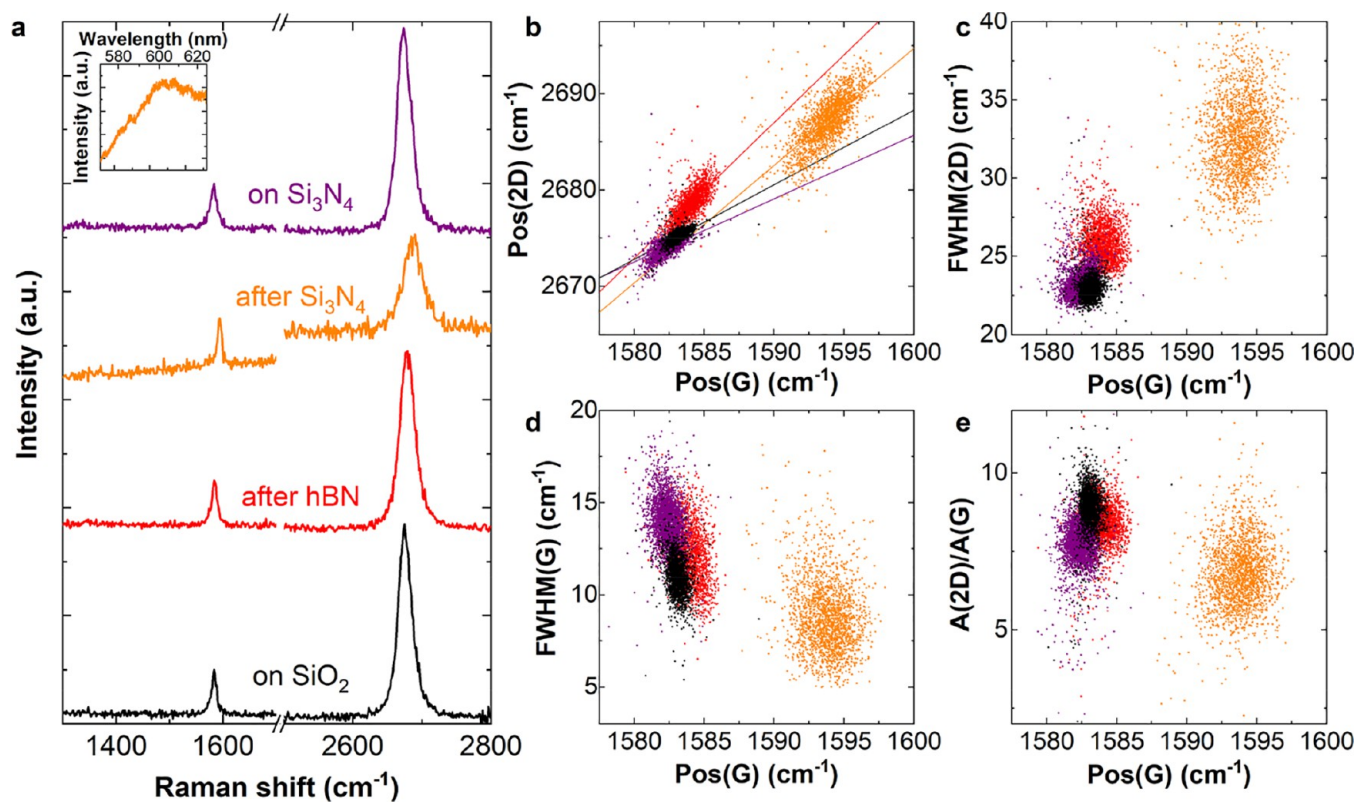

Figure 3. (a) Representative spectra of SLG SCs for the different fabrication steps. Inset: photoluminescence of 1L-hBN after $Si_3N_4$ deposition. (b−e) Pos(2D), FWHM(2D), FWHM(G), $A(2D)/A(G)$ as a function of Pos(G). The color code is the same as in panel a.

The bottom-layer SLG, transferred and after hBN-capping, and top-layer SLG, are within the intrinsic SLG range in terms of doping ($E_F$ < 100 meV).[99,100] After $Si_3N_4$ deposition, the bottom-layer $E_F$ increases to ∼250 meV.[99,100] The linear fit to Pos(2D) as a function of Pos(G) in Figure 3b gives ΔPos(2D)/ΔPos(G) ∼ 0.78, 0.66, 1.41, 1.22 for bottom-layer SLG transferred on $SiO_2$, top-layer on $Si_3N_4$, bottom-layer after hBN capping, and bottom-layer after $Si_3N_4$ deposition, respectively. This indicates the coexistence of strain and doping, modulated during the assembly steps. The presence (or coexistence) of biaxial strain cannot be ruled out. Considering the Grüneisen parameters[102−104] and the unstrained, undoped Pos(G)[93,105] for intrinsic SLG as above, we estimate a mean uniaxial(biaxial) strain ε ∼ 0.07%(∼0.03%) and 0.03% (∼0.01%), for SLG after transferring on $SiO_2$ (bottom-layer) and on $Si_3N_4$ (top-layer), respectively. The bottom SLG after hBN capping has ε ∼ 0.1%(∼0.04%) while, after $Si_3N_4$ deposition, considering doping,[98] ε ∼ 0.13%(∼0.05%).

To monitor the uniformity of the Raman response throughout the fabrication of the DSLGs, we map 48 SLG SCs, 24 bottom-layer (b1−4 arrays), and 24 top-layer (t1−4 arrays), on four different portions of a 150 mm wafer. Figure 4 plots false-color maps of $I(D)/I(G)$, FWHM(2D), FWHM-(G), $A(2D)/A(G)$ for the four assembly stages. Each map is taken with 10 μm steps. At a given stage, the Raman data do not show significant variations between SLG belonging to the same portion of the wafer. The same applies between SLG from different parts. This implies that the spread in points in Figure 3b−e is representative of the variation of the Raman peaks within individual SLG SCs while, over the scale of the entire wafer, SLG SCs have uniform properties. Small (10−20 μm wide) bilayer graphene (BLG) regions form at nucleation seeds during CVD (on 38/48 of the analyzed crystals, see broad 2D peak central pixels in Figure 4b[93]).

$I(D)/I(G)$, Figure 4a, is negligible throughout the fabrication, except for b1−4 after $Si_3N_4$ deposition, where it is within 0.1 (0.25) for 59% (90%) of the crystals (see also the average values in Table 2). FWHM(2D), Figure 4b, progressively increases upon fabrication on b1−4, while it is comparable for b1−4 and t1−4 after transfer on $SiO_2$ and $Si_3N_4$. FWHM(G) and $A(2D)/A(G)$, Figure 4c,d, are comparable for all SLG SCs, except for b1−4 after $Si_3N_4$ deposition, where they decrease due to $E_F$ > 100 meV.

Thus, our wafer-scale Raman characterization reveals that the top-SLG in the DSLG is comparable to micromechanically exfoliated flakes in terms of doping,[99] strain,[108] and strain fluctuations.[109,110] The transfer of hBN has marginal effect on the properties of the bottom-SLG. However, it plays a key role in preserving the structural integrity of the crystals, and avoiding the formation of Raman-active defects during $Si_3N_4$ deposition, thus preventing μ degradation. The Raman analysis shows an increase in doping, strain and strain fluctuations in the bottom SLG after the PECVD process. However, the PECVD process results in an homogeneous dielectric layer, crucial for reproducible operation of DSLG modulators.[34]

We then investigate the electrical transport properties of the transferred SLG-SCs using back-gated multiterminal devices at RT and exposed to air. This allows us to monitor two key performance parameters for SLG integration in a photonic circuit: contact resistance ($R_c$) and μ.

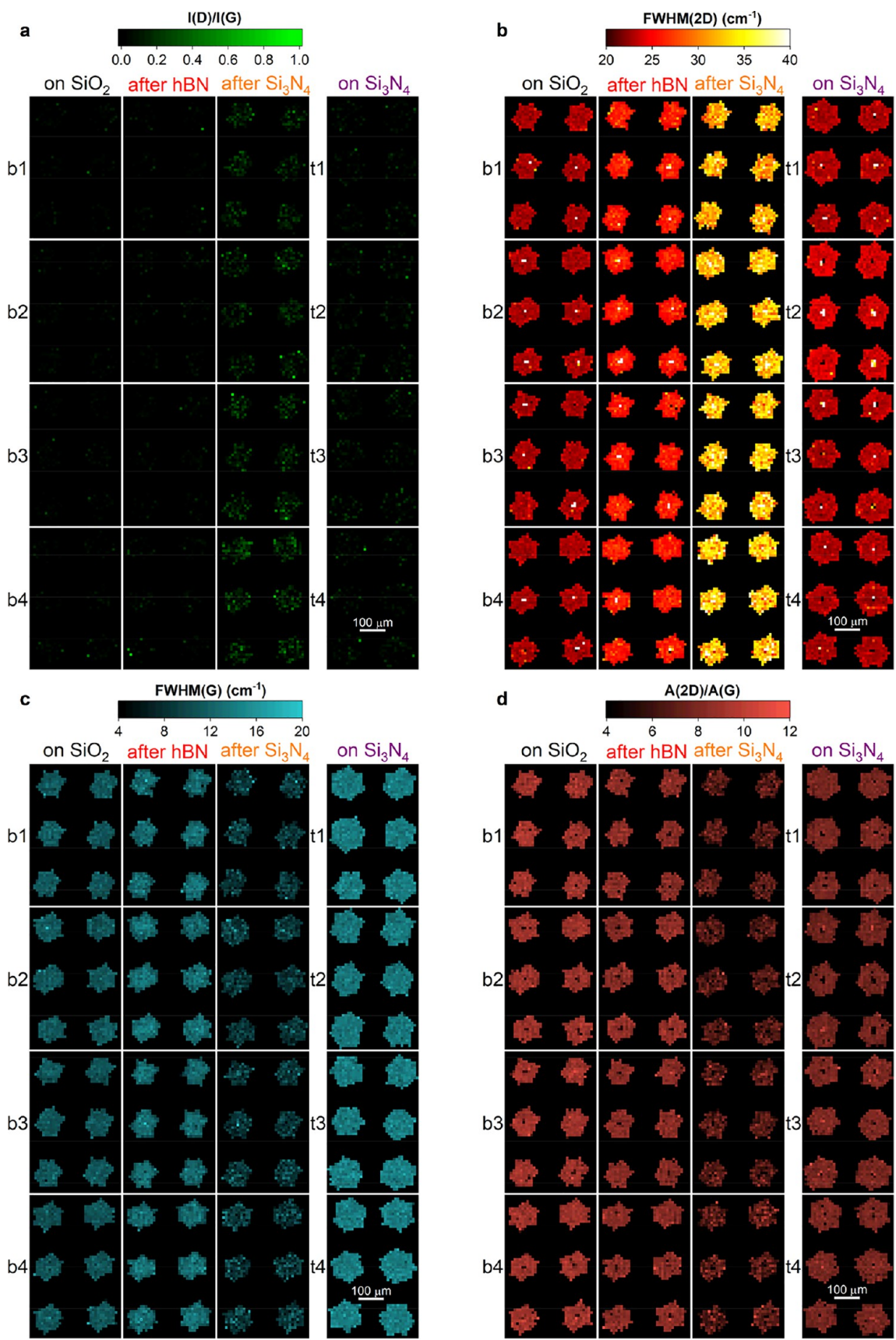

**Figure 4.** Wafer-scale Raman mapping at each fabrication step over different quadrants of the wafer. (a−d) Maps of I(D)/I(G), FWHM(2D), FWHM(G), $A(2D)/A(G)$. Raman mapping is performed at each assembly stage over bottom (b1−4) and top SLG arrays (t1−4).

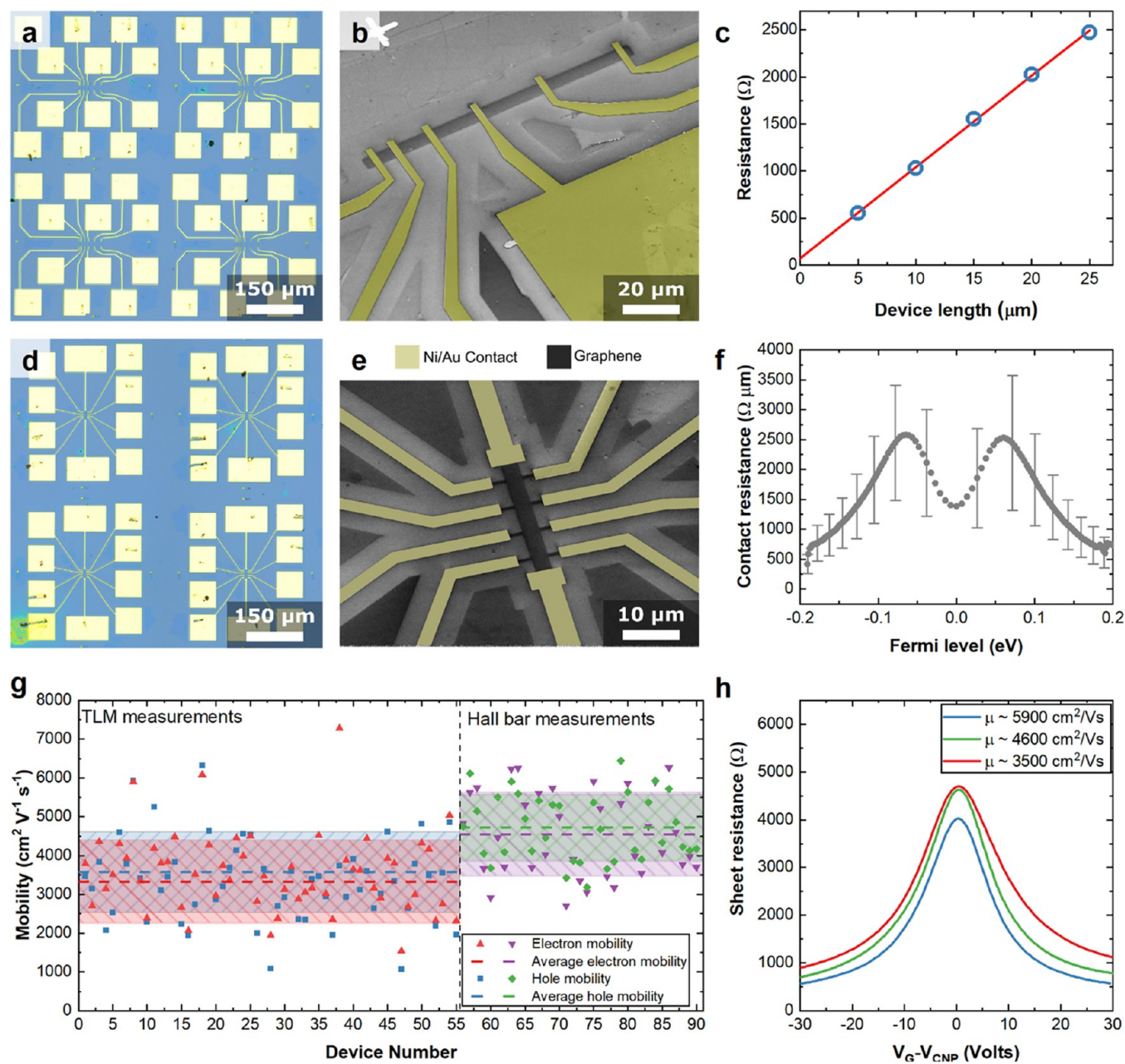

**Figure 5.** Wafer-scale electrical characterization. (a) Optical micrograph of TLM structures. (b) SEM image of representative TLM structure. (c) Estimation of $R_C$ via linear fit of TLM measurements. (d) Optical micrograph of Hall bars. (e) SEM image of representative Hall bar. (f) $R_C$ as a function of $E_F$. (g) Statistics of e and h mobility from TLM and Hall measurements. Dashed lines represent the average $\mu$. Shaded areas indicate the standard deviation. (h) Representative field effect curves for 3 Hall bars with $\mu$ ~3500, ~4600, ~5900 cm$^2$ V$^{-1}$ s$^{-1}$.

To quantify $R_c$, we use transfer-length method (TLM)[111] devices, as in Figure 5a,b, defined by EBL, reactive-ion etching and thermal evaporation of metallic contacts. Ni/Au 7/60 nm top contacts evaporated <10$^{-5}$ mbar provide the highest performing configuration in terms of yield (>80% of working devices) and Rc when compared to Cr, Ti, and Ni and to other contact geometries, such as one-dimensional side contacts.[112]

By measuring the two-terminal resistance over different channel lengths ($l$) we extrapolate the residual resistance at $l$ = 0, which corresponds to 2 × $R_c$,[111] Figure 5c. This procedure can be repeated for different $E_F$, set by the back-gate voltage ($V_G$), to obtain $R_c$ as a function of $E_F$, as for Figure 5f, showing the statistical average over 56 devices and error bars as standard deviations. $R_c$ remains <2500 $\Omega$ $\mu$m in the neutrality region and is ~500 $\Omega$ $\mu$m for $E_F$ > 0.2 eV, required in the operation of modulators at telecom wavelengths.[2] The SLG $E_F$ must be set at energies larger than half of the photon energy in order to work at the edge of Pauli blocking.[42,43,113] At 1550 nm the photon energy is 0.8 eV, so that $E_F$ must be set slightly above 0.4 eV.[34] These $R_c$ are comparable to those previously reported for ultrahigh $\mu$ > 10$^5$ cm$^2$ V$^{-1}$ s$^{-1}$ devices.[112] We get $\mu$ from 56 TLM structures as well as 36 Hall bars, in Figure 5d,f. The SLG resistivity, $\rho$, for the TLM devices is obtained from a linear fit of TLM channels (Figure 5c) as a function of $V_G$. The Hall bar $\rho$ is derived from four-terminal measurements and fitted as for ref 114. In Figure 5g, dashed lines indicate the average $\mu$ for both e and h, whereas the shaded areas represent the standard deviation. The average $\mu$ from Hall bars (~4750 $cm^2\ V^{-1}\ s^{-1}$ for h and ∼4600 $cm^2\ V^{-1}\ s^{-1}$ for e) is higher than TLM (∼3600 and ∼3350 $cm^2\ V^{-1}\ s^{-1}$, respectively). This could be caused by two factors. (1) For each TLM, $\rho$ is estimated from an average of 5 channels, with a total length of 75 $\mu$m, whereas the channel length in a Hall bar is 8 $\mu$m, comparable to that used in typical SLG transport measurements.[114] (2) Parasitic doping by the contacts has an effect in two-terminal TLM measurements,[115,116] not present in four-terminal Hall bar measuremnts.[117] Figure 5h plots 3 representative traces of $\rho$ as a function of $V_G$, from Hall bars with high (∼5900 $cm^2\ V^{-1}\ s^{-1}$), low (∼3500 $cm^2\ V^{-1}\ s^{-1}$), and average (∼4700 $cm^2\ V^{-1}\ s^{-1}$) $\mu$.

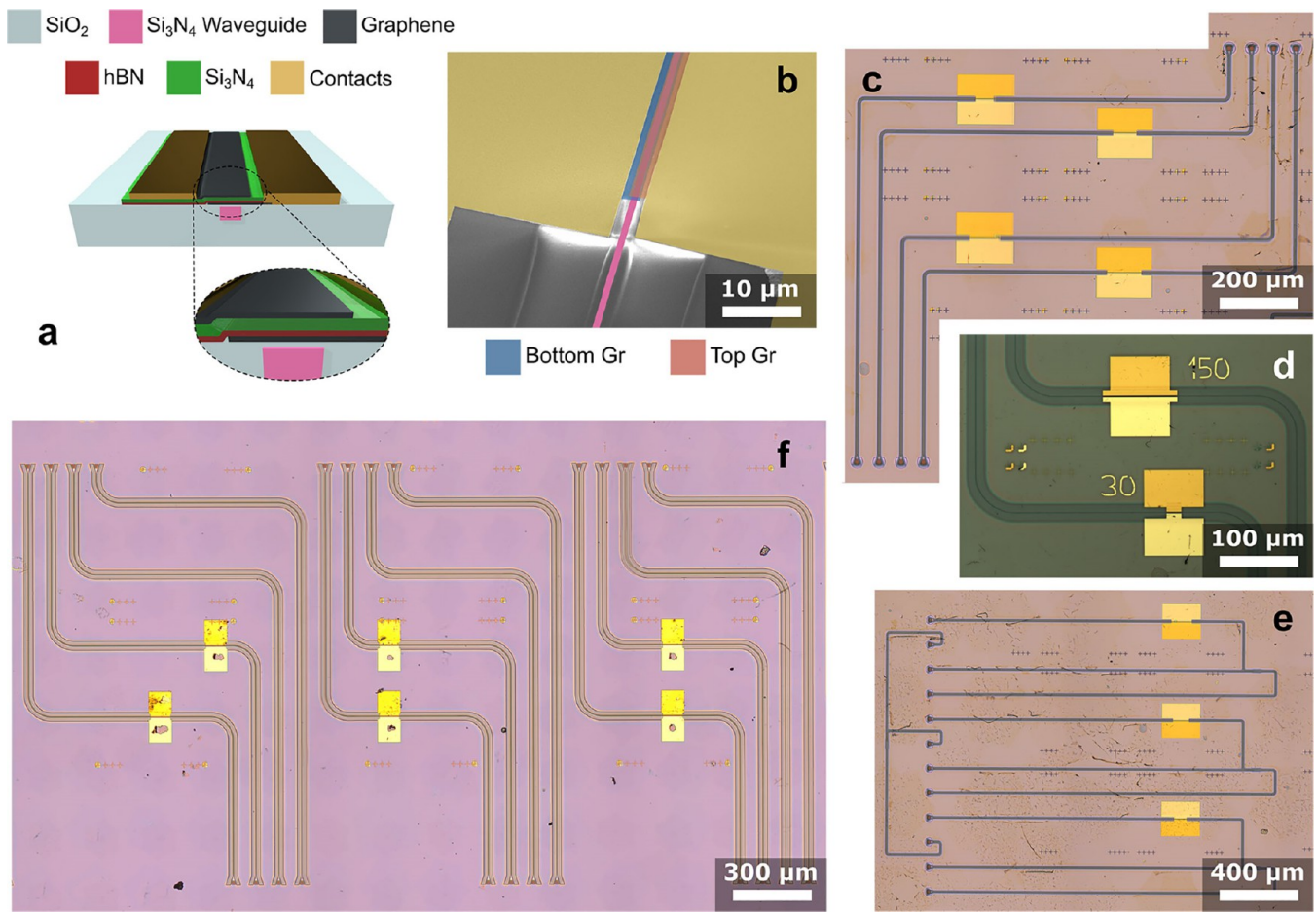

Figure 6. (a) Cross-section of DSLG EAMs. The $Si_3N_4$ WG core is 1500 nm wide and 260 nm thick, the buried oxide is 15 $\mu$m, and the distance of the metal electrodes from the WG edge is 700 nm. (b) SEM image of DSLG EAM showing the overlap of the two SLG (blue and red) above the photonics WG (pink). (c−f) Optical micrographs of four chips with DSLG EAMs.

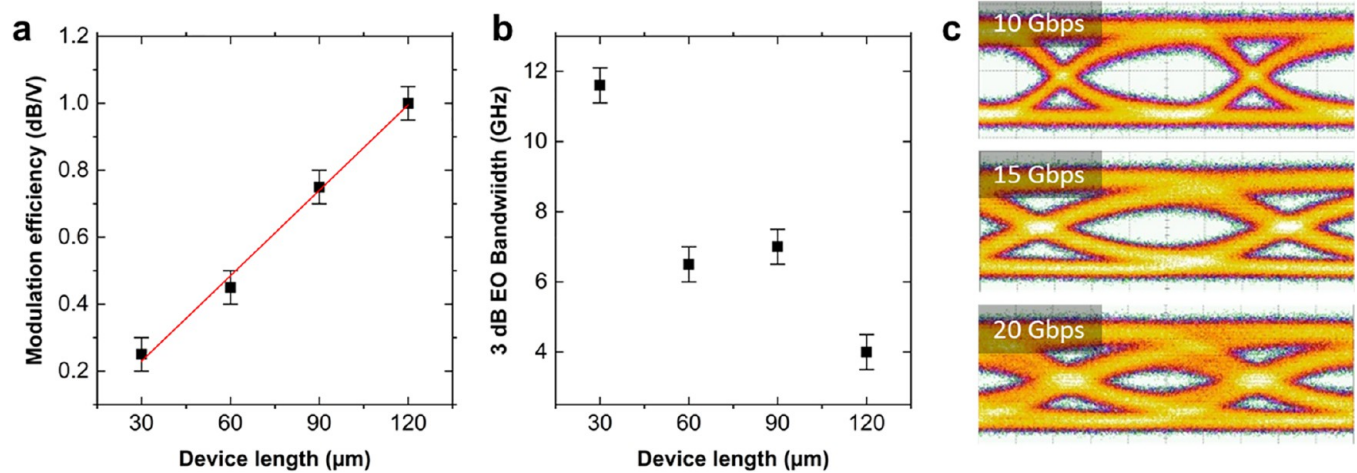

Figure 7. DSLG EAM characterization. (a) Modulation efficiency as a function of device length. (b) 3 dB EO BW as a function of devices length (c) Eye diagrams at 10, 15, and 20 Gbps.

EAMs are based on the modulation of the surface optical conductivity at optical frequencies induced by electric field effect.[41,118] SLG absorption is changed by moving $E_F$ above the Pauli blocking condition.[42,43,113] This can be done by applying gating in a capacitor-like structure, with SLG used as one or both capacitor plates.[2] In our DSLG geometry, a reciprocal self-gating is obtained with $V_G$, resulting in modulation of the surface carrier density, i.e., electro-absorption.[34] The main advantages of this approach are the larger electro-absorption effect, due to the presence of two SLG, approximately twice that of SLG,[34] and the possibility to use undoped WGs, enabling integration onto any already existing platform, such as SOI for SiPh or $Si_3N_4$ on Si.[34]

Here we use a 150 mm $Si_3N_4$ photonic platform, with 260 nm $Si_3N_4$ on a 15 $\mu$m buried $SiO_2$. The 1500 nm wide WG is designed to support a transverse-electric field (quasi-TE) mode at 1550 nm.[17] The top cladding is thinned to ~40 nm to maximize the evanescent coupling of the optical mode with the DSLG stack. The core of the modulators is the DSLG capacitor, comprising a SLG/hBN/$Si_3N_4$/SLG stack. The cross-section and a SEM image of a representative device is in Figure 6a,b (see the Methods for details). We prepare 30 SLG/hBN/$Si_3N_4$/SLG stacks on 30 WGs to fabricate 30 EAMs with different lengths (Figure 6c–f). This allows us to benchmark the reproducibility of the fabrication process at wafer scale through optoelectronic characterization of the devices.

We test key performance parameters: static (DC-biased) and dynamic (DC-biased + RF) modulation depth, electro-optical (EO) BW, and eye diagram opening. We characterize the EAMs in static and dynamic (i.e., driven by a time varying electrical signal) mode and collect the data to perform a statistical study of performance, Figure 7. We first consider the transmission as a function of $V_G$. Modulation is obtained by tuning $E_F$ of both SLG layers from complete optical absorption ($E_F$ < 0.4 eV at 1550 nm) toward transparency ($E_F$ > 0.4 eV).[34] The static characterization on wafer scale shows modulation efficiency ~0.25, 0.45, 0.75, 1 dB $V^{-1}$ for ~30, 60, 90, 120 $\mu$m EAMs, respectively, Figure 7a. We then characterize the EO BW, i.e., the BW of the conversion efficiency, defined as the ratio between the output and the input power,[17] from the electrical signal driving the modulator and the optical modulated signal at the output of the modulator.[17] This parameter determines the maximum operating speed and is typically affected by $R_C$.[119] The EAM BW is mainly limited by its RC time constant,[119] i.e., the series resistance ($R$) of the device multiplied by the DSLG capacitance, $C$, given by the series of gate dielectric capacitance and quantum capacitance of the two SLGs,[120] with $R = R_C + R_S$ of the SLG section between DSLG capacitor and metal contacts. As $C$ is proportional to the device length, while $R$ is inversely proportional to it, we would expect a length-independent 3 dB electro-optical BW. However, Figure 7b shows that the BW changes with length, with longer devices having lower BW. We obtain ~11.5, 6.5, 7.4 GHz for 30, 60, 120 $\mu$m, respectively. The reason is that a further contribution to $R$ comes from the output 50 $\Omega$ impedance of the vector network analyzer (VNA) used to perform the measurements (see the Methods). This is the main limiting resistive contribution because of our low $R_c$ ~ 500 $\Omega$ $\mu$m at $E_F$ > 0.2 eV.

We then test the DSLG EAMs using a non-return-to-zero (NRZ) electrical driving signal,[58] i.e., a digital two-level sequence, generated with a pattern generator (PG) (Anritzu MP1800A). This instrument allows us to obtain pseudorandom binary sequences (PRBS), i.e., deterministic binary sequences of bits with statistical behavior similar to a pure random sequence,[27] with adjustable lengths (up to $2^{31}$-1 bits). The signal is applied to the DSLG EAMs electrodes through a RF cable and a bias-tee. This generates a modulated optical signal, detected by a high-frequency (70 GHz) photodetector (Finisar XPDV3120) connected to a sampling digital oscilloscope (Infinium DCA 83484A, BW ~ 50 GHz). By doing so, we can visualize on the oscilloscope the resulting eye diagram,[121] Figure 7c. This gives the frequency dependent ER and 3 dB EO BW as a function of device length, and 10/15/20 Gbps data-rate.[121] The eye diagram measurement of the data stream along with ER and 3 dB EO BW demonstrate EAM at 20 Gbps on wafer scale. Our wafer-scale fabrication approach may also be used on different photonic platforms, e.g., SOI. The smaller WG cross section, 480 nm × 220 nm, would reduce the modulator stack capacitance, thus improving EAM speed.

The change from $Si_3N_4$ to SOI, as reported in ref 47, increases the EO BW to at least 30 GHz, and the data rate to 50 Gbps in a 100 $\mu$m EAM. Improving the SLG quality, in terms of $\mu$ after $Si_3N_4$ encapsulation, can increase performance in terms of insertion loss per unit length. Assuming a maximum absorption ~0.1 and <0.001 dB $\mu m^{-1}$ in the transparency region for $\mu$ > 3000 $cm^2$ $V^{-1}$ $s^{-1}$ at 0.4 eV, the EAM length can be reduced to 50 $\mu$m, with a maximum ER = 5 dB and a halved capacitance. By reducing the RC constant, we expect to approximately double its BW with respect to the 100 $\mu$m device, thus achieving ~60 GHz. This optimization, combined with a SOI WG, could result in EAMs competitive with present microring based SOI modulators[28,122] and SiGe EAMs.[29] The added value of SLG-based EAMs is the broad operation spectrum, from O (1300 nm) to L-band (>1625 nm) and beyond, while SiGe modulators are restricted to the C band (1530–1565 nm),[123] and Si microring modulators are limited to resonant wavelengths.[124]

## CONCLUSIONS

We presented the full process flow (from growth, to transfer, integration on WGs, and photonic devices fabrication) for SLG-based photonics on wafer-scale. Our approach yields high-quality uniform SLG on wafer-scale, as indicated by statistical spectroscopic and electrical characterizations. We used wafer scale hBN encapsulation to minimize damage during dielectric deposition. We applied this to realize double SLG electro-absorption modulators on the passive $Si_3N_4$ platform. Our approach is easier and more reproducible, in terms of yield and uniformity, compared to the transfer of a continuous SLG film over the full wafer area, because it is based on individual crystal matrices. SLG single crystals have higher mobility than polycrystalline films, with high-quality top contacts, with a reproducible contact resistance ~500 $\Omega$ $\mu$m. Our approach can be used for other photonics building blocks, such as photodetectors and mixers, as well as for resonant structures, including microrings for modulation, switching and filtering, and nonresonant ones, like interferometers.

## METHODS

SLG crystal matrices are grown on 25 $\mu$m Cu foils (Alfa Aesar no. 46365). Prior to SLG growth, each foil is electropolished in an electrolyte consisting of water, ethanol, phosphoric acid, isopropyl alcohol, and urea, as for ref 79. The Cu foil is patterned using UV lithography. Cu is spin-coated with a Shipley S1813 positive photoresist, baked at 110 °C for 1 min, and exposed to UV light using a Cr mask containing the required seeding pattern (UV dose

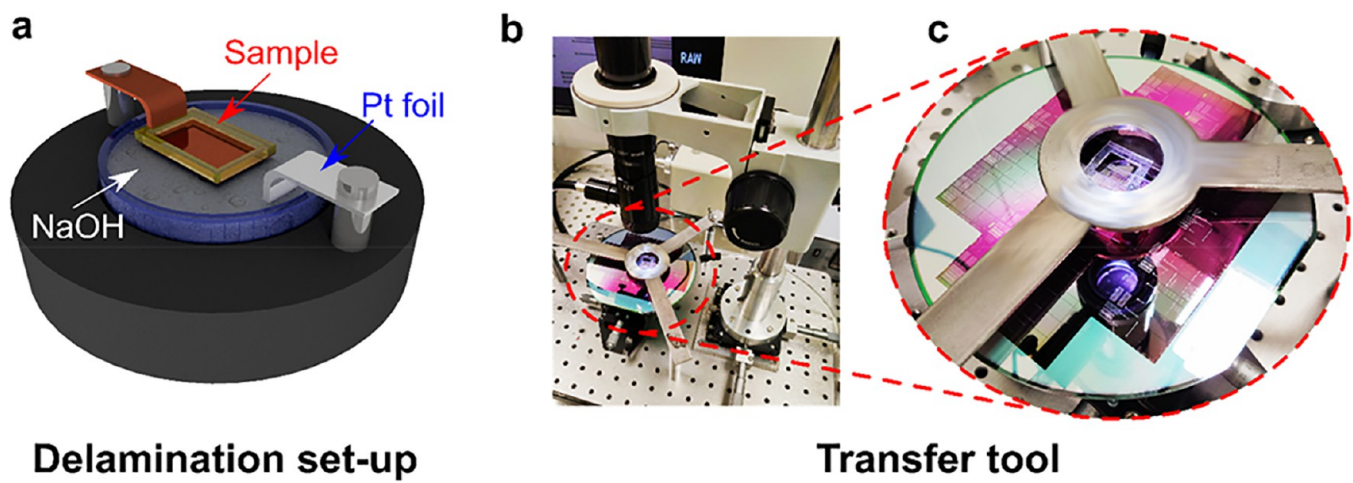

Figure 8. (a) Schematic electrochemical delamination setup. (b) Transfer tool holding the delaminated SLG sample and a 150 mm $Si_3N_4$ wafer patterned with the photonic WG circuits of Figure 6. (c) Close-up of SLG/PMMA membrane, with a PDMS frame aligned onto the target wafer.

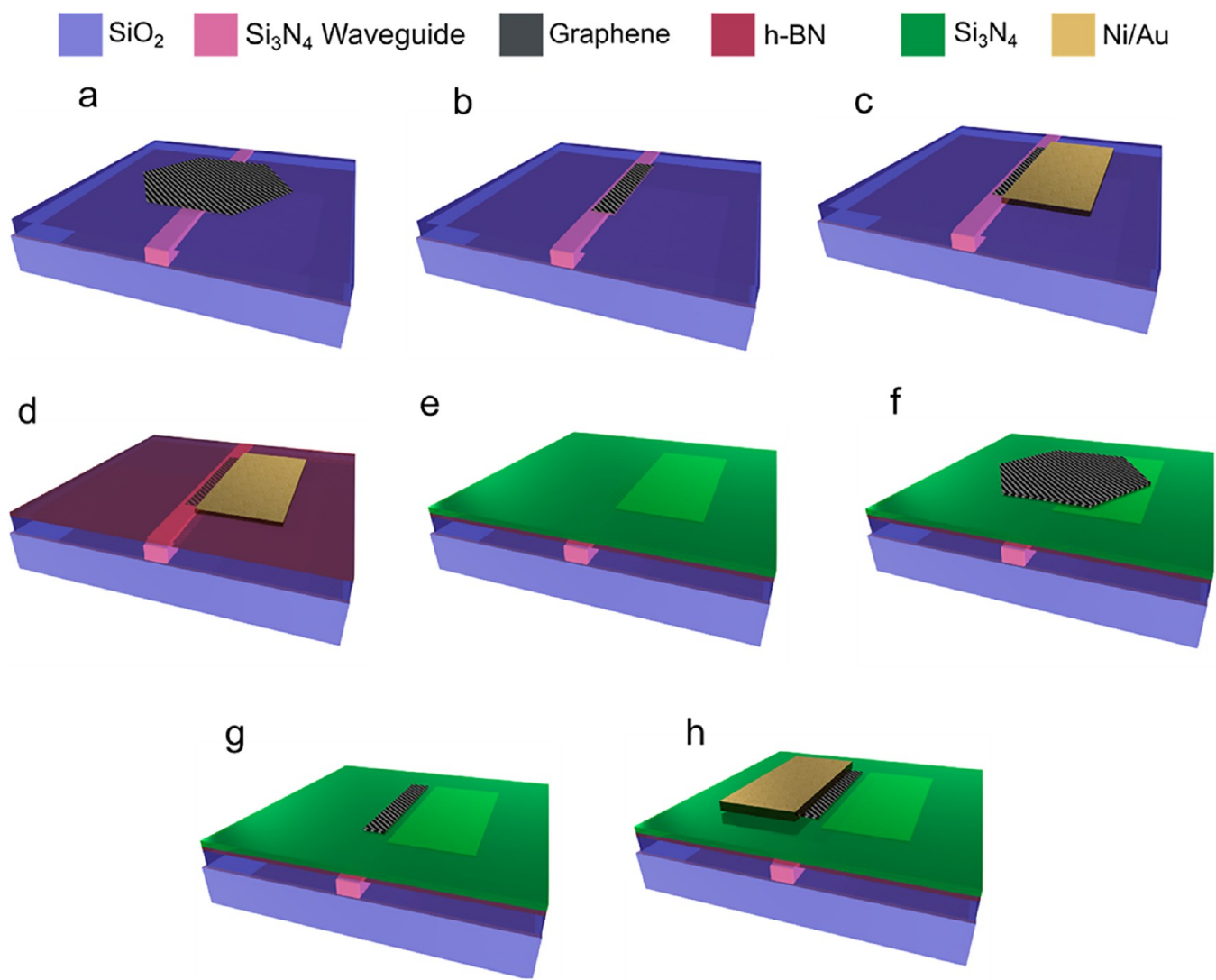

Figure 9. Process flow for DSLG EAM fabrication. (a) SC SLG transfer on WG. (b) SLG patterning using EBL and RIE. (c) Ni/Au contacts deposition using evaporation and lift-off. (d) 1L-hBN transfer on top. (e) $Si_3N_4$ deposition by PECVD. (f) Top layer SLG SC transfer. (g) Top SLG patterning. (h) Top contact deposition.

∼200 mJ cm$^{-2}$). Twenty-five nanometer Cr is thermally evaporated (Sistec) at $1 \times 10^{-5}$ mbar, followed by lift-off in acetone. The samples are then rinsed in isopropyl alcohol. Growth is performed in an Aixtron BM Pro cold-wall reactor at 25 mbar and 1060 °C. The samples are kept under Ar flow during the T ramp-up, and are annealed for 10 min at the growth T. Growth is performed by flowing 0.5 sccm $CH_4$, 50 sccm $H_2$ and 900 sccm Ar. Following the 20 min growth, heating is switched off and the sample is cooled to <120 °C under Ar flow.

SLG on Cu is then coated with a support polymer (100 nm PMMA 950 K and 1.5 μm PPC) and a PDMS frame is attached to the perimeter of the Cu foil. SLG electrochemical delamination is performed in 1 M NaOH. Cu/SLG is used as the anode, and ∼2.4 V is applied with respect to a Pt counter electrode, Figure 8a. The voltage is adjusted to maintain a current ∼3 mA to avoid excessive formation of $H_2$ bubbles, which may cause damage to SLG. The freestanding polymer/SLG membrane is then removed from the electrolyte, rinsed 3 times in DI water, then dried in air.

The lamination of SLG on the target wafer is performed in a transfer tool, shown in Figure 8b, with a close-up of the SLG/PMMA membrane with a PDMS frame aligned onto the target wafer in Figure 8c. The target wafer is placed on a micrometric stage with three-axis translational and azimuthal rotational movement, Figure 8b. Alignment of the WGs to the SLG SC matrix is performed exploiting the SLG contrast on the polymer membrane in transmission mode, Figure 1d. The optical system of the transfer tool consists of a 0.58−7× microscope objective with coaxial illumination, and a DSLR camera with a 2× adapter tube, giving a final magnification ∼1.16−14×.

Following alignment, the wafer is heated to 100 °C using the inbuilt stage heater with a proportional-integral-derivative (PID) controller, and the membrane is brought into contact with the wafer to laminate the SLG. Heating the wafer reduces the adhesion of PDMS, and the frame can be then detached from the wafer, Figure 7b. Depending on the geometry of the wafer, several cycles of the above procedure are performed to populate the wafer with SLG SCs. For a typical SLG SC matrix of $25 \times 40$ mm$^2$, 16 cycles populate 90% of a 150 mm wafer. Finally, the wafer is placed in acetone to remove the support polymer, followed by a rinse in isopropyl alcohol.

The fabrication of the DSLG modulator stack is performed as follows. A matrix of SLG SCs is transferred on the target wafer and aligned to the $Si_3N_4$ WG, Figure 9a. The bottom layer SLG is spin-coated with PMMA 950 A4 (Microchem), patterned using EBL and etched using RIE, Figure 9b. Contacts to the bottom SLG are fabricated using EBL and thermal evaporation of 7 nm Ni and 60 nm Au, followed by lift-off in acetone, Figure 9c. A $2 \times 2.5$ cm$^2$ polycrystalline 1L-hBN (Graphene Laboratories, Inc.) grown on Cu foil via CVD[125] is then electrochemically delaminated from Cu and transferred on the chips of the wafer via semidry transfer.[76] $Si_3N_4$ (17 nm) is deposited using PECVD at 350 °C, Figure 9e. The top layer of the modulator is fabricated following the same protocol of transfer (Figure 9f), etching (Figure 9g), and contacting (Figure 9h).

## AUTHOR INFORMATION

**Corresponding Authors**
- **Camilla Coletti** − *Center for Nanotechnology Innovation @ NEST - Istituto Italiano di Tecnologia, I-56127 Pisa, Italy; Graphene Labs, Istituto Italiano di Tecnologia, 16163 Genova, Italy;* orcid.org/0000-0002-8134-7633; Email: camilla.coletti@iit.it
- **Marco Romagnoli** − *Photonic Networks and Technologies Lab, CNIT, 56124 Pisa, Italy; INPHOTEC, 56124 Pisa, Italy; CamGraPhiC, 56124 Pisa, Italy;* Email: mromagnoli@cnit.it

**Authors**
- **Marco A. Giambra** − *Photonic Networks and Technologies Lab, CNIT, 56124 Pisa, Italy; INPHOTEC, 56124 Pisa, Italy; Center for Nanotechnology Innovation @NEST - Istituto Italiano di Tecnologia, I-56127 Pisa, Italy;* orcid.org/0000-0002-1566-2395
- **Vaidotas Mišeikis** − *Center for Nanotechnology Innovation @ NEST - Istituto Italiano di Tecnologia, I-56127 Pisa, Italy; Graphene Labs, Istituto Italiano di Tecnologia, 16163 Genova, Italy;* orcid.org/0000-0001-6263-4250
- **Sergio Pezzini** − *Center for Nanotechnology Innovation @ NEST - Istituto Italiano di Tecnologia, I-56127 Pisa, Italy; Graphene Labs, Istituto Italiano di Tecnologia, 16163 Genova, Italy; NEST, Scuola Normale Superiore and Istituto Nanoscienze-CNR, I-56127 Pisa, Italy;* orcid.org/0000-0003-4289-907X
- **Simone Marconi** − *Photonic Networks and Technologies Lab, Tecip Institute, Scuola Superiore Sant'Anna, 56124 Pisa, Italy*
- **Alberto Montanaro** − *Photonic Networks and Technologies Lab, CNIT, 56124 Pisa, Italy*
- **Filippo Fabbri** − *Center for Nanotechnology Innovation @ NEST - Istituto Italiano di Tecnologia, I-56127 Pisa, Italy; Graphene Labs, Istituto Italiano di Tecnologia, 16163 Genova, Italy; NEST, Scuola Normale Superiore and Istituto Nanoscienze-CNR, I-56127 Pisa, Italy;* orcid.org/0000-0003-1142-0441
- **Vito Sorianello** − *Photonic Networks and Technologies Lab, CNIT, 56124 Pisa, Italy*
- **Andrea C. Ferrari** − *Cambridge Graphene Centre, Cambridge University, Cambridge, U.K.;* orcid.org/0000-0003-0907-9993

Complete contact information is available at:
https://pubs.acs.org/10.1021/acsnano.0c09758

**Author Contributions**
†M.A.G. and V.M. contributed equally to this work.

**Author Contributions**
‡C.C. and M.R. are joint last coauthors.

**Notes**
The authors declare no competing financial interest.

## ACKNOWLEDGMENTS

We acknowledge funding from the European Union H2020 Graphene Flagship under grant agreements nos. 785219 and 881603, ERC grants Hetero2D, GSYNCOR, EPSRC grants EP/L016087/1, EP/K01711X/1, EP/K017144/1, EP/N010345/1.